\documentclass[conference]{IEEEtran}
\IEEEoverridecommandlockouts
\usepackage{cite}
\usepackage{amsmath,amssymb,amsfonts}
\usepackage{algorithmic}
\usepackage{graphicx}
\usepackage{textcomp}
\usepackage{epsfig}
\usepackage{multirow}
\usepackage{float}
\usepackage{subfigure}
\def\BibTeX{{\rm B\kern-.05em{\sc i\kern-.025em b}\kern-.08em
		T\kern-.1667em\lower.7ex\hbox{E}\kern-.125emX}}
\begin{document}
	
	\title{Learning Constellation Map with Deep CNN for Accurate Modulation Recognition\\
				\thanks{This research was financially supported by National Research Foundation of Korea (NRF) through Creativity Challenge Research-based Project (2019R1I1A1A01063781), and in part by the Priority Research Centers Program through the NRF funded by the Ministry of Education, Science and Technology (2018R1A6A1A03024003).}
	}
	
	\author{Van-Sang Doan$^{\ast }$, Thien Huynh-The$^{\ast }$, Cam-Hao Hua$^{\dagger}$, Quoc-Viet Pham$^{\ddagger }$, and Dong-Seong Kim$^{\ast }$\\
		$^{\ast }$ICT Convergence Research Center, Kumoh National Institute of Technology, Korea\\
		$^{\dagger}$Department of Computer Science and Engineering, Kyung Hee University, Korea\\
		$^{\ddagger}$Research Institute of Computer, Information and Communication, Pusan National University, Korea\\
		 Email: \{vansang.doan,thienht,dskim\}@kumoh.ac.kr,~hao.hua@oslab.khu.ac.kr,~vietpq@pusan.ac.kr}

	\maketitle

\begin{abstract}
Modulation classification, recognized as the intermediate step between signal detection and demodulation, is widely deployed in several modern wireless communication systems.
Although many approaches have been studied in the last decades for identifying the modulation format of an incoming signal, they often reveal the obstacle of learning radio characteristics for most traditional machine learning algorithms.
To overcome this drawback, we propose an accurate modulation classification method by exploiting deep learning for being compatible with constellation diagram.
Particularly, a convolutional neural network is developed for proficiently learning the most relevant radio characteristics of gray-scale constellation image.
The deep network is specified by multiple processing blocks, where several grouped and asymmetric convolutional layers in each block are organized by a flow-in-flow structure for feature enrichment.
These blocks are connected via skip-connection to prevent the vanishing gradient problem while effectively preserving the information identify throughout the network.
Regarding several intensive simulations on the constellation image dataset of eight digital modulations, the proposed deep network achieves the remarkable classification accuracy of approximately $87\%$ at 0~dB signal-to-noise ratio (SNR) under a multipath Rayleigh fading channel and further outperforms some state-of-the-art deep models of constellation-based modulation classification.

\end{abstract}


\section{Introduction}

Nowadays, regarding the accelerated development of wireless communications, many innovative communication standards and technologies have been released to satisfy numerous high-quality services~\cite{PhamIoT,Zhang2019}.
Consequently, spectrum monitoring becomes more challenging due to the disordered radio-frequency signals in a dense radio environment.
Automatic signal recognition and modulation classification can improve the performance of intelligent spectrum analysis which plays a great functionality in modern wireless communication systems like fifth-generation (5G)~\cite{HuynhComLet,HuynhWCNC,PhamSurvey}.

Modulation classification, which allows blindly identifying the modulation fashion of an incoming radio signal at the receiver, is fundamentally regarded to as a multi-class decision making task.
Accurately classifying advanced modulations under harmful transmission conditions, such as multipath fading and additive noise, remains an open topic which is attracting much interest from signal processing and communication communities.
Numerous modulation classification have been introduced in the last decades for civil and military applications (e.g., radio fault detection, interference signal identification, and dynamic spectrum access), which can be grouped into two categorized: likelihood-based (in which unknown parameters of a classification model are estimated by expectation/conditional maximization algorithm~\cite{Hameed-2009}) and feature-based (in which the signal characteristics are extracted by feature engineering techniques~\cite{Huang-2017}).
Despite exploiting several advanced machine learning (ML) algorithms, modeling a modulation classifier from the raw in-phase (I) and quadrature (Q) data of radio signal suffers from poor performance at low signal-to-noise ratio (SNR)~\cite{Yang-2019}.

\begin{figure}[!t]
	\centering
	\includegraphics[width=8.75cm]{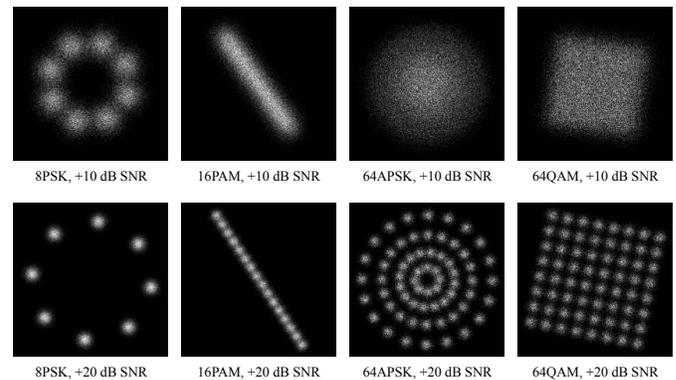}
	\caption{Constellation images of some common modulation formats at different SNR levels.}
	\label{fig_constellation}
\end{figure}

Constellation diagram of radio signal has been widely used for digital modulation design and analysis~\cite{Jajoo-2019,Kumar-2020,FWang-2017}.
Among many favorable characteristics, constellation diagram is able to provide the insightful visualization of signal structure and the relative correlation between various modulation states via the position and distribution of constellation points~\cite{Kumar-2020}.
Many approaches have been studied for constellation-based modulation classification by incorporating advanced image processing and machine learning (ML) algorithms, but they almost fail in effectively learning noisy constellation patterns~\cite{FWang-2017} (see some examples shown in Fig.~\ref{fig_constellation}).

As an innovative branch of ML, deep learning (DL) has achieved the outstanding success of performance in many informatics fields, including signal processing for communication, thanks to two key benefits: automatic learning features from high-dimensional unstructured data and effectively dealing with big data.
To overcome the weak discrimination of traditional ML algorithms for modeling visual-based constellation patterns, we take advantage of a DL~\cite{LeCun-2015} to capture high-level features from constellation images for the performance improvement of modulation classification.
Concretely, we propose a novel deep convolutional neural network (CNN) for being compatible with the input data of constellation image, in which the network architecture involves several processing modules for comprehensively learning relevant information in multi-scale feature representations.
Every module has two convolutional blocks associated via skip-connection to maintain information identity, where each block is specified by a flow-in-flow structure of multiple grouped and asymmetric convolutional layers gain rich features.
For performance evaluation, we generate a challenging dataset of gray-scale constellation images from the modulated signals of eight digital modulation formats, in which bivariate histogram and exponential decay mechanisms are applied to enhance the visualization of constellation points.
Based on several intensive simulation results, the designed network is robust with different sizes of constellation image and outperforms other existing CNN-based modulation classification approaches in terms of accuracy while maintaining a reasonable computational complexity.

\section{Related Work}

The topic of digital modulation classification using the visual information of constellation shape has been early investigated by Mobasseri~\cite{Mobasseri-2000}, in which a fuzzy c-means clustering algorithm is studied to identify the unknown constellation map of a modulation signal.
The statistical features achieved by clustering pixels in a constellation image are learned by a naive Bayes classifier that is able to predict the modulation of a radio signal in the inference process.
To effectively deal with the scatter plot of received symbols in a two-dimensional (2-D) distribution, Pedzisz and Mansour~\cite{Pedzisz-2005} rotates the constellation map for better analysis of the in-phase component. 
Accordingly, the Fourier series expansion of the high-order cumulant is then extracted for model learning by a feed-forward neural network.
The approach is evaluated on an effortless datasetof three modulation formats, including BPSK, QPSK, and 8APSK, where the classification rate decreases along the increment of modulation order.

Recently, with the great performance in terms of accuracy for various image processing and computer vision tasks~\cite{HuynhTII-2020, HuynhINS-2020, Hao-2019, Hao-2018}, DL has been studied for communication to overcome some obstacles of traditional machine learning algorithms, in which several CNN architectures~\cite{Gonzalez-2018} are developed for learning pattern recognition model of modulation from constellation maps~\cite{Wang-2019}.
For example, Wang et al.~\cite{Wang-2017} proposed a CNN-based deep learning method for intelligent constellation diagram analysis, wherein a simple architecture is designed with two convolutional layers alternately connected with pooling layers for six-modulation-format classification.
Based on simulation results, the network reports high accuracy at the SNR larger than 20 dB.
In~\cite{Zhang-2018}, the constellation map of a modulation signal is pre-processed with power normalization and three-dimensional (3D) stokes space mapping to visual enrichment.
For training a modulation classifier, the transfer learning technique is performed via MobileNetV2~\cite{Sandler-2018}, a backbone network for image classification, to take over the rich feature representation. 
Peng et al.~\cite{Peng-2019} converted constellation diagram to three-channel color image with a visualization enhancement scheme, wherein the pixel value is determined via the measurement distance between constellation points and the centroid of pixel.
Subsequently, the modulation classification method is benchmarked with different CNN backbones, such as AlexNet and GoogletNet, for performance analysis in comparison with some conventional machine learning algorithms.
The combination of regular constellation image and contrast enhanced grid constellation image is also considered by Huang et al.~\cite{Huang-2019} for improving the accuracy of high-order modulations like 64QAM.
Concretely, the deep feature maps extracted from two constellation images using two parallel convolutional flows are fused at an intermediate level for learning a classification model with fully connected layers.
Besides constellation diagram, the image of spectrum features is further taken into account for modulation classification.
Zeng et al.~\cite{Zeng-2019} exploited a time-frequency analysis technique to visualize the spectrum of signal frequency, in which the spectrogram of a modulation signal is obtained using the short-time Fourier transform with a Hann window of length of 40 samples.
The spectrogram images are then learned by a conventional CNN, where the network architecture is specified by four convolutional networks for extract high-level features.

\section{Deep Convolutional Neural Network for Modulation Classification}
\subsection{Dataset Generation: Signal Model Configuration}

Nowadays, modern communication systems rapidly replace analog modulations by digital ones thanks to a better compatibility with digital data in computing machines and a stronger immunity against interference.
In this work, four principal modulation classes are considered with the following mathematical description.
\begin{itemize}
	\item Phase-shift keying (PSK)
	\begin{equation} s\left ( t \right ) = A \mathrm {cos}\left ( 2\pi f_c t + \pi u\left ( t \right ) \right ), \end{equation} 
	where $u\left ( t \right )$, as the informative symbol in time, is modulated by the sinusoidal carrier signal specified by the amplitude $A$ and the frequency $f_c$.
	\item Pulse amplitude modulation (PAM)
	\begin{equation} s\left ( t \right ) = u\left ( t \right ) g\left ( t \right ) A \mathrm {cos}\left ( 2\pi f_c t \right ), \end{equation}
	where $g\left ( t \right )$ is the pulse-shaping factor which is determined as follows
	\begin{equation}
	g\left ( t \right )=\mathrm{sinc}\left (  \frac{t}{T} \right ) \frac{\mathrm {cos}\left ( \pi \alpha t/T \right )}{1-4 \alpha^2 t^2/T},
	\end{equation} 
	where $T$ is the pulse width and $\alpha$ refers to as the roll-off factor between 0 and 1.
	\item Quadrature amplitude modulation (QAM)
	\begin{equation}
	\begin{aligned}
	s\left ( t \right ) & =  \left | u\left ( t \right ) \right |\mathrm {cos}\left ( \mathrm {arg}\left \{ u\left ( t \right ) \right \} \right ) \mathrm{cos} \left ( 2 \pi f_c t \right ) \\
	& - \left | u\left ( t \right ) \right |\mathrm {sin}\left ( \mathrm {arg}\left \{ u\left ( t \right ) \right \} \right ) \mathrm{sin} \left ( 2 \pi f_c t \right ),
	\end{aligned}
	\end{equation}  
	where $\left | u\left ( t \right ) \right |$ and $\mathrm {arg} \left \{ u\left ( t \right ) \right\}$ are the amplitude and phase of complex baseband signal, respectively.
	\item Amplitude-phase shift keying (APSK)
	\begin{equation} s\left ( t \right ) = u\left ( t \right ) \mathrm {cos}\left ( 2\pi f_c t \right ). \end{equation} 
	$u\left ( t \right )$ is determined as follows:
	 \begin{equation}
	\begin{aligned}
	u\left ( t \right )&=\left \{ \begin{array}{cc}
	r_1 e^{j\left ( \frac{2\pi}{n_1}k \right )}, \mathrm{for~} k=0,1,\dots,n_1-1,\\ 
	r_2 e^{j\left ( \frac{2\pi}{n_2}k \right )}, \mathrm{for~} k=0,1,\dots,n_2-1,\\ 
	\vdots \\
	r_N e^{j\left ( \frac{2\pi}{n_N}k \right )}, \mathrm{for~} k=0,1,\dots,n_N-1,\\ 
	\end{array} \right. \\
	\end{aligned}
	\end{equation} where $r_1,\dots,r_N$ refers to as the levels of amplitudes and $n_i$ is the order of PSK modulations.
\end{itemize}
In particular, eight digital modulation formats are taken into account, including 16QAM, 64QAM, 4PAM, 16PAM, QPSK, 8PSK, 16APSK, and 64APSK, in which some high-order ones, such as 64QAM and 64APSK in Fig.~\ref{fig_constellation}, can be really challenging for accurate classification in low SNR.

\begin{figure*} [!t]
	\centering
	\subfigure[]
	{	\includegraphics[width=3.8in]{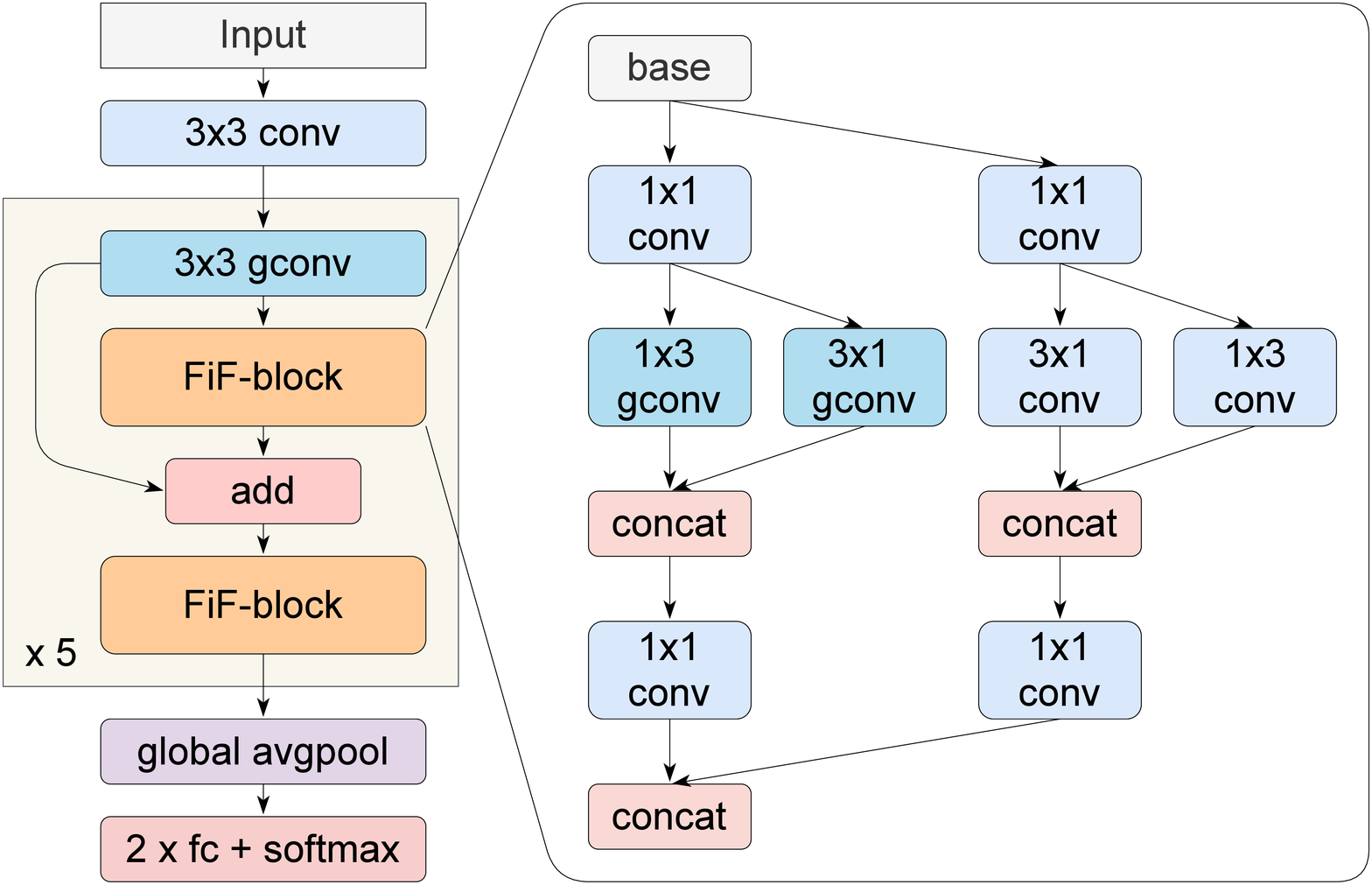}
		\label{architecture}
	}
	\subfigure[]
	{	\includegraphics[width=2.584in]{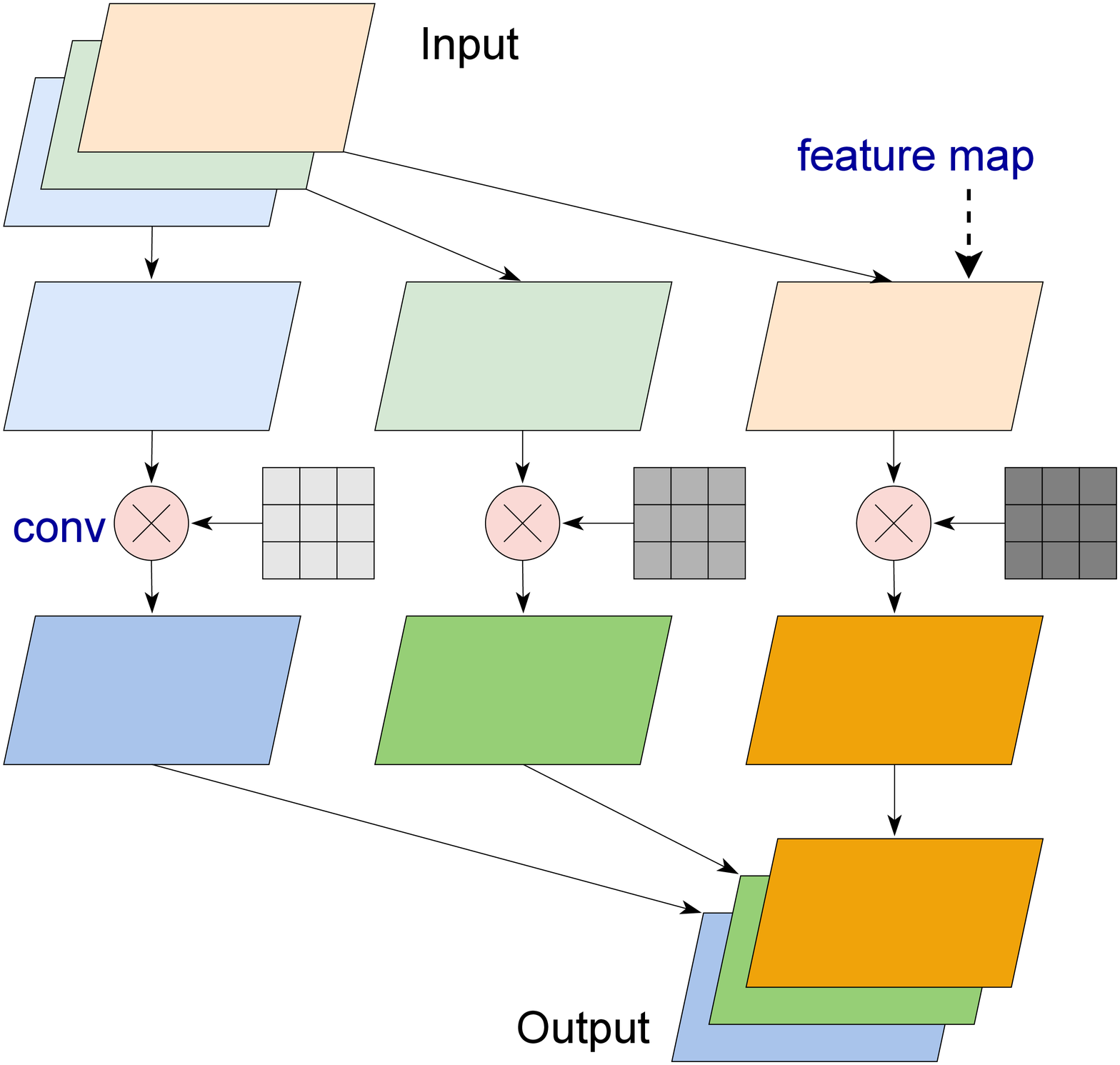}
		\label{channelwise}
	}
	\caption{Illustration: (a) FiF-Net: the overall network architecture consists of five primary modules for extracting deep features at multi-scale maps. Each feature extraction module has two FiF-blocks which are constituted by various regular convolutional ($\mathtt{conv}$) and grouped convolutional ($\mathtt{gconv}$) layers of asymmetric kernels, and (b) channel-wise convolution: each spatial map in a feature volume is associated with only one spatial kernel.}
	\label{network}
\end{figure*}

%\begin{figure*}[!t]
%	\centering
%	\begin{subfigure}[b]{0.5\textwidth}
%		\includegraphics[width=\textwidth]{architecture.eps}
%		\caption{FiF-Net}
%		\label{architecture}
%	\end{subfigure}
%	\hspace{8mm}
%	\begin{subfigure}[b]{0.34\textwidth}
%		\includegraphics[width=\textwidth]{channelwise.eps}
%		\caption{Channel-wise convolution}
%		\label{channelwise}
%	\end{subfigure}
%	\caption{Illustration: (a) FiF-Net: the overall network architecture consists of five primary modules for extracting deep features at multi-scale maps. Each feature extraction module has two FiF-blocks which are constituted by various regular convolutional ($\mathtt{conv}$) and grouped convolutional ($\mathtt{gconv}$) layers of asymmetric kernels, and (b) channel-wise convolution: each spatial map in a feature volume is associated with only one spatial kernel.}
%	\label{network}
%\end{figure*}

Constellation diagram, fundamentally referred to as the two-dimensional (2-D) presentation of a modulation signal by drawing samples as scattered points on a complex plane (aka I/Q plane), is usually exploited for automatic signal analysis thanks to its perceptive information of modulation type and order.
Indeed, the constellation (or cluster) number indicates the modulation order, hence the modulation type can be recognized via the relative pattern of constellation distribution in the plane. 
Regularly, the constellation of a low-order modulation signal is perhaps distinguished in a comfortable transmission condition (e.g., the constellation map of 8PSK at $+20$ dB SNR in Fig.~\ref{fig_constellation}). 
On the contrary, recognizing the constellation of a high-order modulation signal is nearly impossible under a synthetic channel deterioration (e.g., the constellation diagrams of 64APSK and 64QAM at $+10$ dB SNR), wherein the radio characteristics are strongly modified by such impairments as multipath fading and additive white Gaussian noise (AWGN).
Therefore, the constellation-based modulation classification remains a challenging task to properly discriminate high-order modulations under strong channel impairments.
%Many approaches have been studied for this topic by incorporating advanced image processing and machine learning algorithms, but they almost fail in effectively learning noisy constellation patterns.

For evaluating the performance of classification model, we introduce a challenging dataset of constellation image from modulation signal under a synthetic channel impairment with multipath Rayleigh fading channel (i.e., following the ITU Pedestrian A channel profile with the propagation path delays in range $\left \{0, 110, 190, 410  \right \}$ ns, the average path gains in range $\left \{0, -9.7, -19.2 -22.8  \right \}$ dB) and AWGN, where SNR varies from $-20$ dB to $+30$ dB with a step of $5$ dB. 
For each modulation at every SNR, we independently generate 4000 constellation images of spatial size $200 \times 200$.
As a result, the dataset has 352,000 images in total.
Due to the constraint of visualizing constellation diagram on small-size image, many scattered points are allocated within a pixel coordinate without care of the power and position of signal samples.
To handle these problems, we leverage a bivariate histogram and an exponential decay mechanism to obtain gray-scale constellation image, where the pixel value is inferred as follows
\begin{equation}
V_j=\frac{1}{K}\sum_{i=1}^{K}P_{ij}e^{-\mu d_{ij}},
\end{equation} 
where $P_{ij}=I^2_{ij}+Q^2_{ij}$ is the power of signal sample point $i$-th in pixel $j$, $d_{ij}$ is the distance between the point $i$-th and the center of pixel $j$, and $\mu$ as an exponential decay rate is set to $0.5$ in this work. 

\subsection{FiF-Net: Convolutional Neural Network for Learning Constellation Map}

In this section, we introduce a novel deep network, denoted FiF-Net, for learning modulation patterns from constellation image, where the network architecture is shown in Fig.~\ref{architecture}.
At the beginning of network, an input layer configured by the size of $200 \times 200 \times 1$ to be compatible with the volume size of constellation image is followed by a convolutional layer with 64 kernels of size $3 \times 3$ and a rectified linear unit (ReLU) layer to acquire coarse features.
Afterwards, the network consists of five primary modules for deeply mining the fine features at multi-scale representational maps.
Each module has two sophisticated convolutional blocks, called Flow-in-Flow block (FiF-block), which are cascaded along the network backbone.
For details, the first element of the module is a grouped convolutional layer with 64 groups of one kernel of size $3 \times 3$.
This layer performs channel-wise separable convolution (also known as depth-wise separable convolution), where its operating mechanism is illustrated in Fig.~\ref{channelwise}.
The second element is FiF-block with the detailed structure zoomed in Fig.~\ref{architecture}, wherein there are two main processing flows: the first flow is principally specified by two grouped convolutional layers arranged in parallel, meanwhile the second flow is configured by regular convolutional layers with the same structure of the first flow.
Instead of adopting the two-dimension (2-D) kernel of size $3 \times 3$, we utilize two 1-D asymmetric kernels of size $1 \times 3$ and $3 \times 1$ for both regular and grouped convolutional layers to reduce the number of trainable weights and preserve an equivalent learning efficiency.
The output feature maps from these convolutional layers are then merged in the depth dimension via depth-wise concatenation layer ($\mathtt{concat}$).
Two unit convolutional layers with the kernel of size $1 \times 1$ is to re-scale the depth size of output feature volume resulted by the concatenation operation.
This process can be generally expressed as follows
\begin{equation}
y_{\mathtt {FiF-block}}^{H \times W \times D}=\mathtt{concat} \left \{ y_{\mathtt {gconv-flow}}^{H \times W \times D_1}, y_{\mathtt {conv-flow}}^{H \times W \times D_2} \right \},\\
\end{equation} 
where $y_{\mathtt {FiF-block}}$ is the output of FiF-block with the volume size of $H \times W \times D$.
The depth size $D = D_1 + D_2$ is the sum of those of two flows $y_{\mathtt {gconv-flow}}$ and $y_{\mathtt {conv-flow}}$ determined as follows
\begin{equation}
\begin{aligned}
y_{\mathtt {gconv-flow}}^{H \times W \times D_1}&=\mathtt{conv}_{1 \times 1} \left \{ \mathtt{concat} \left \{ y_{\mathtt {gconv} 1\times 3}^{H \times W \times d_{11}}, y_{\mathtt {gconv} 3\times 1}^{H \times W \times d_{12}} \right \} \right \}, \\[3pt]
y_{\mathtt {conv-flow}}^{H \times W \times D_2}&=\mathtt{conv}_{1 \times 1} \left \{ \mathtt{concat} \left \{ y_{\mathtt {conv} 1\times 3}^{H \times W \times d_{21}}, y_{\mathtt {conv} 3\times 1}^{H \times W \times d_{22}} \right \} \right \},
\end{aligned}
\end{equation} 
where $y_{\mathtt {gconv} 1\times 3}$, $y_{\mathtt {gconv} 3\times 1}$, $y_{\mathtt {conv} 1\times 3}$, and $y_{\mathtt {conv} 3\times 1}$ are the outputs of regular convolutional and grouped convolutional layers, respectively.
It is important to notice that, in FiF-blocks, a regular convolutional layer is followed by a ReLU layer, whereas a grouped convolutional layer is tailed by a clipped ReLU layer to prevent the output from becoming too large.
\begin{equation}
\begin{aligned}
\mathtt{ReLU}\left ( x \right )&=\left \{ \begin{array}{ll}
0, & x<0,\\ 
x, & x\geq 0,
\end{array} \right. \\[3pt]
\mathtt{ClippedReLU}\left ( x \right )&=\left \{ \begin{array}{ll}
0, & x<0,\\ 
x, & 0 \leq  x \leqslant m,\\
m, & x \geq m,
\end{array} \right.
\end{aligned}
\end{equation}
where $m$ refers to as the ceiling value for input clipping.
To prevent the vanishing gradient issue caused by activation functions in FiF-block and maintain the informative identity of previous layers, skip-connection is deployed via an element-wise addition layer ($\mathtt{add}$) as follows
\begin{equation}
y_{\mathtt {skip-connection}}=\mathtt{add} \left \{ x_{\mathtt {FiF-block}}, y_{\mathtt {FiF-block}} \right \} 
\end{equation}
where the input of FiF-block $x_{\mathtt {FiF-block}}$ is the output of the $3 \times 3$ grouped convolutional layer.
The module is finalized with the second FiF-block.
By following this architecture, the spatial size of output feature volume halves for every module.

The network is finalized with a global average pooling layer ($\mathtt{avgpool}$) with the pool size of $7 \times 7$, two fully connected layers ($\mathtt{fc}$) (where the number of hidden nodes in the second $\mathtt{fc}$ layer is identical to the number of modulation formats considered for classification), and a softmax layer.
The detailed configurations of FiF-Net are given in Table~\ref{tab_configuration}.
 
\section{Performance Evaluation}
\subsection{Model Robustness}

\begin{table}[!t]
	\centering
	\footnotesize
	\caption{Detailed configuration of network architecture.}
	\setlength{\tabcolsep}{6pt}
	\begin{tabular}{|l|l|}
		\hline
		\textbf{Component} &\textbf{Detailed description} \\
		\hline
		$\mathtt{input}$ 	&  $\mathtt{constellation~image}$ \\ \hline
		$\mathtt{conv}$ 	&  $64~\mathtt{conv}~3\times3,\mathtt{~stride}~\left (1,1  \right )$ \\\hline			
		$5\times\mathtt{module}$ & $5 \times \left \{ \begin{array}{ll}
		64~\mathtt{gconv}~3\times3,\mathtt{~stride}~\left (2,2  \right )\\ 
		\mathtt{block} : \left \{ \begin{array}{ll}
		4\times32~\mathtt{conv}~1\times1\\ 
		1\times32~\mathtt{conv}~1\times3\\ 
		1\times32~\mathtt{conv}~3\times1\\ 
		1\times32~\mathtt{gconv}~1\times3\\ 
		1\times32~\mathtt{gconv}~3\times1\\ 
		3\times\mathtt{concatenation}\\ 
		\end{array} \right \}\\ 
		\mathtt{element-wise~addition}\\ 
		\mathtt{block} : \left \{ \begin{array}{ll}
		4\times32~\mathtt{conv}~1\times1\\ 
		1\times32~\mathtt{conv}~1\times3\\ 
		1\times32~\mathtt{conv}~3\times1\\ 
		1\times32~\mathtt{gconv}~1\times3\\ 
		1\times32~\mathtt{gconv}~3\times1\\ 
		3\times\mathtt{concatenation}\\ 
		\end{array} \right \}\\ 
		\end{array} \right \}$ \\ \hline 
		$\mathtt{avgpool}$  & $\mathtt{global~average~pooling~7\times7}$\\ 
		$\mathtt{fc1}$  & $\mathtt{128~hidden~nodes}$\\
		$\mathtt{fc2}$  & $\mathtt{16~modulation~formats}$\\ \hline
	\end{tabular}
	\label{tab_configuration}
\end{table}

\begin{figure}[!t]
	\centering
	\subfigure[]
	{	\includegraphics[width=2.7in]{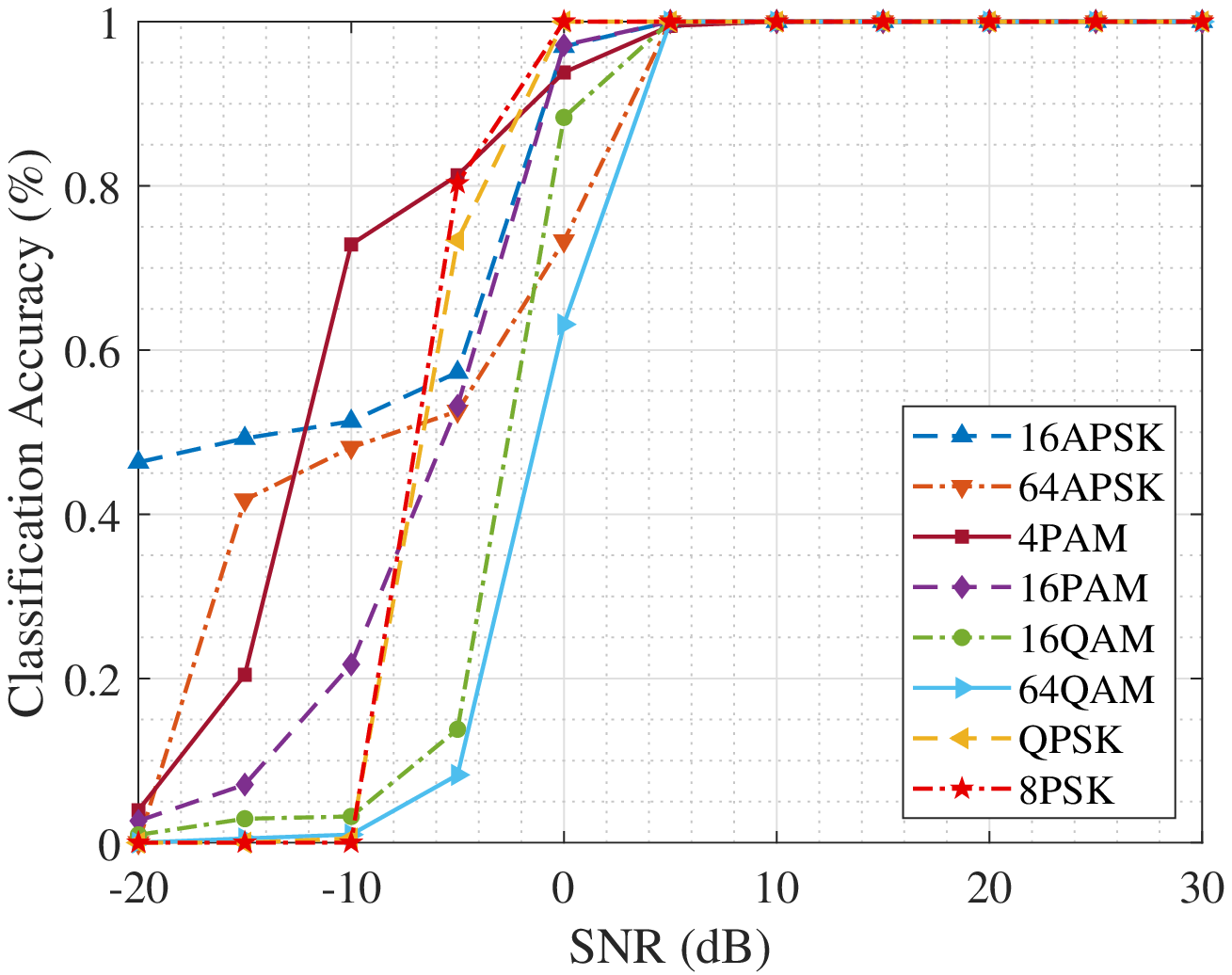}
		\label{fig_accuracy}
	}
	\subfigure[]
	{	\includegraphics[width=2.7in]{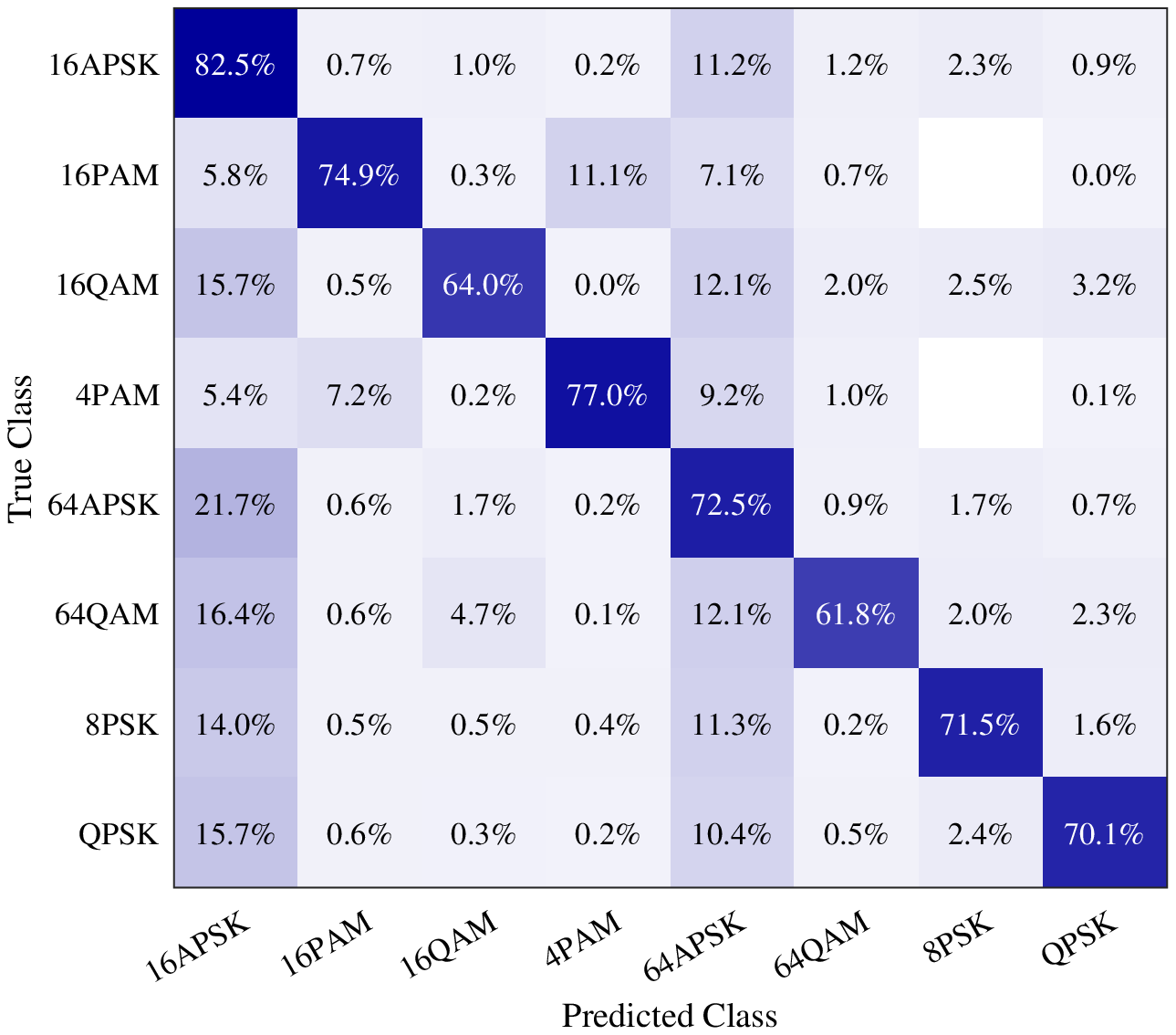}
		\label{fig_confusion}
	}
	\caption{Classification performance of FiF on the testing set: (a) accuracy of eight modulations and (b) confusion matrix of eight-modulation classification at all SNR levels with the overall accuracy is over $72\%$.}
	\label{fig_FiFaccuracy}
\end{figure}

%\begin{figure}[!t]
%	\centering
%	%\includegraphics[width=8cm]{detail-accuracy.eps}
%	\includegraphics[width=8cm]{detail-accuracy.eps}
%	\caption{Performance of eight digital modulations.}
%	\label{fig_accuracy}
%\end{figure}

In the first simulation, we report the classification accuracy of eight modulation formats under the aforementioned synthetic channel impairment, where the numerical results achieved by FiF-Net are plotted in Fig.~\ref{fig_accuracy}.
In general, the classification rate increases along the increment of SNR levels.
Interestingly, many modulations are perfectly classified at $+5$ dB SNR, meanwhile, some of them incredibly achieve the accuracy over $80\%$ at $-5$ dB SNR, for example, 8PSK with $80.31\%$ and 4PAM with $81.28\%$.
It is observed that the classification rate keeps getting worse along increasing the modulation order, for instance, 16APSK is better than 64APSK by approximately $23.7\%$ and 16QAM is more superior than 64QAM by around $25.2\%$ at 0 dB SNR.
In a higher-order modulation, the distance between scattered points distributed in a constellation map is more narrow.
Obviously, a wireless communication system can achieve a higher transmission rate with a higher-order modulation, but the recognition will be less accurate because of the vulnerability of closer constellation points.
For an insightful analysis, we further provide the confusion matrix of eight-modulation classification in Fig.~\ref{fig_confusion}.
Several modulations are mostly confused with 16APSK and 64APSK due to the harmful effect of additive noise and multipath fading channel, in which the amplitude and phase of modulation signals are extremely modified.
For example, 64QAM which reports the worst accuracy of $61.8\%$ is highly confused with 16APSK by $16.4\%$ and 64APSK by $12.1\%$, that means, many 64QAM signals are misidentified as the APSK-modulation signals.
Notably, despite achieving the greatest accuracy of $82.5\%$, 16APSK strongly suffers the misclassification with 64APSK.

%\begin{figure}[!t]
%	\centering
%	\includegraphics[width=8cm]{confusionAllsnr.eps}
%	\caption{Confusion matrix of eight-modulation classification at all SNR levels, where the overall accuracy is over $72\%$}
%	\label{fig_confusion}
%\end{figure}

\subsection{Ablation Study}

In the second simulation, we study the performance of FiF-Net by varying the size (or referred to as image resolution) of generated constellation image in $\left \{ 25\times25, 50 \times 50, 100 \times 100, 200 \times 200 \right \}$, where the overall results are presented in Fig.~\ref{fig_imagesize}.
FiF-Net achieves a greater accuracy of 8-modulation classification further as the image size increases.
With respect to a larger size of constellation images, the scattered points are visualized more explicitly, that makes the deep network can learn more relevant information for better classification.
Besides, the significant accuracy improvement is found at low SNR levels, for example, when doubling the image size from $25\times25$ to $50 \times 50$, the accuracy is improved by approximately $3.5\%$ at $-10$ dB SNR and $24.0\%$ at $0$ dB SNR.
Statistically, with SNR $\leq 0$ dB, FiF-Net can enhance the average classification rate by around $4.9\%$ for each time of doubling the image size.
Furthermore, the network successfully reaches the very high accuracy of around $94.00\%$ at $+5$ dB SNR with the $25\times25$ constellation images.
It should be noted that a greater performance can be obtained by increasing the image size of constellation image for a better quality of visualization, but the computational cost requires more expensively for performing convolution operation on larger feature maps besides much memory consumption for image repository.

\subsection{Method Comparison}

\begin{figure}[!t]
	\centering
	\subfigure[]
	{	\includegraphics[width=2.8in]{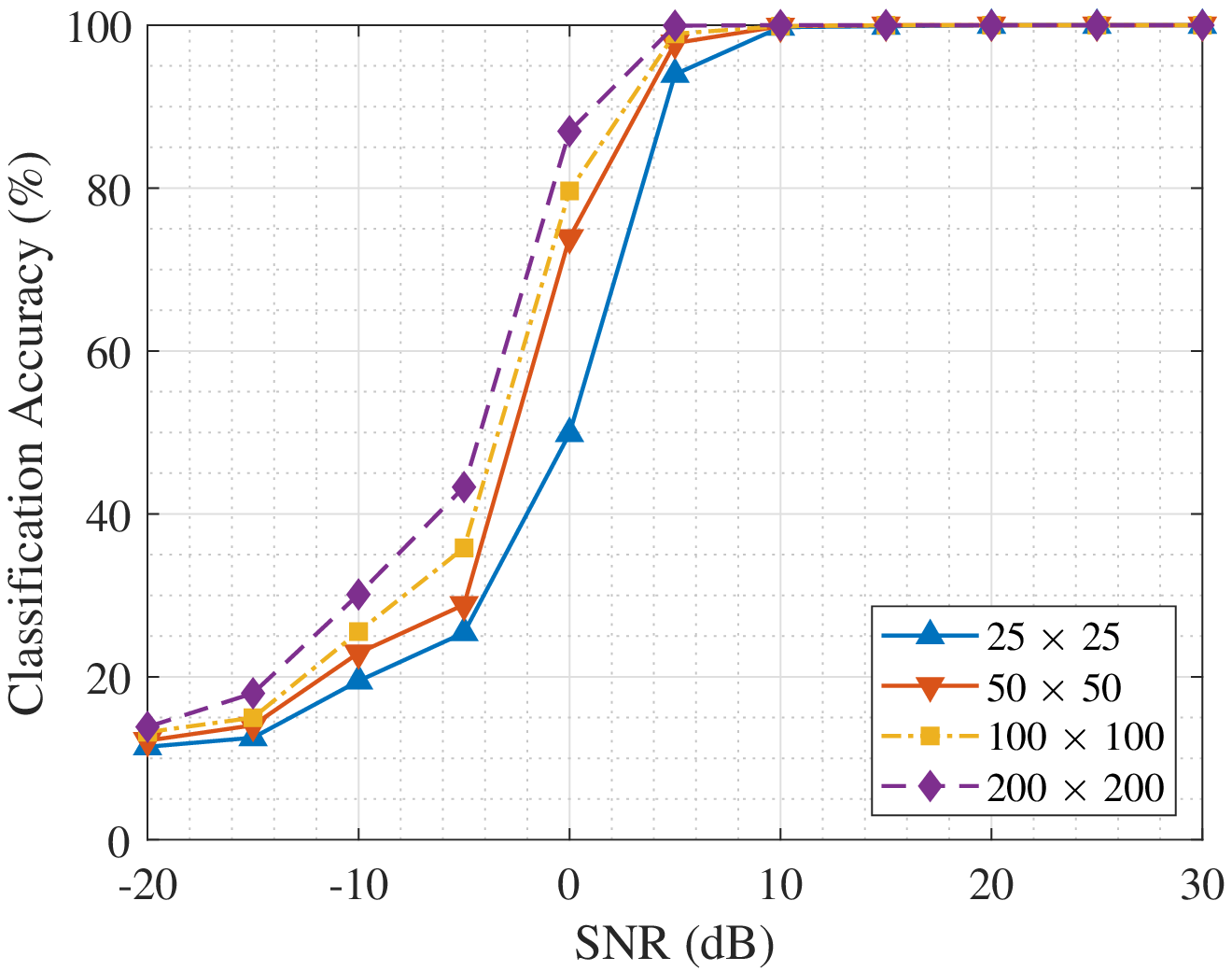}
		\label{fig_imagesize}
	}
	\subfigure[]
	{	\includegraphics[width=2.8in]{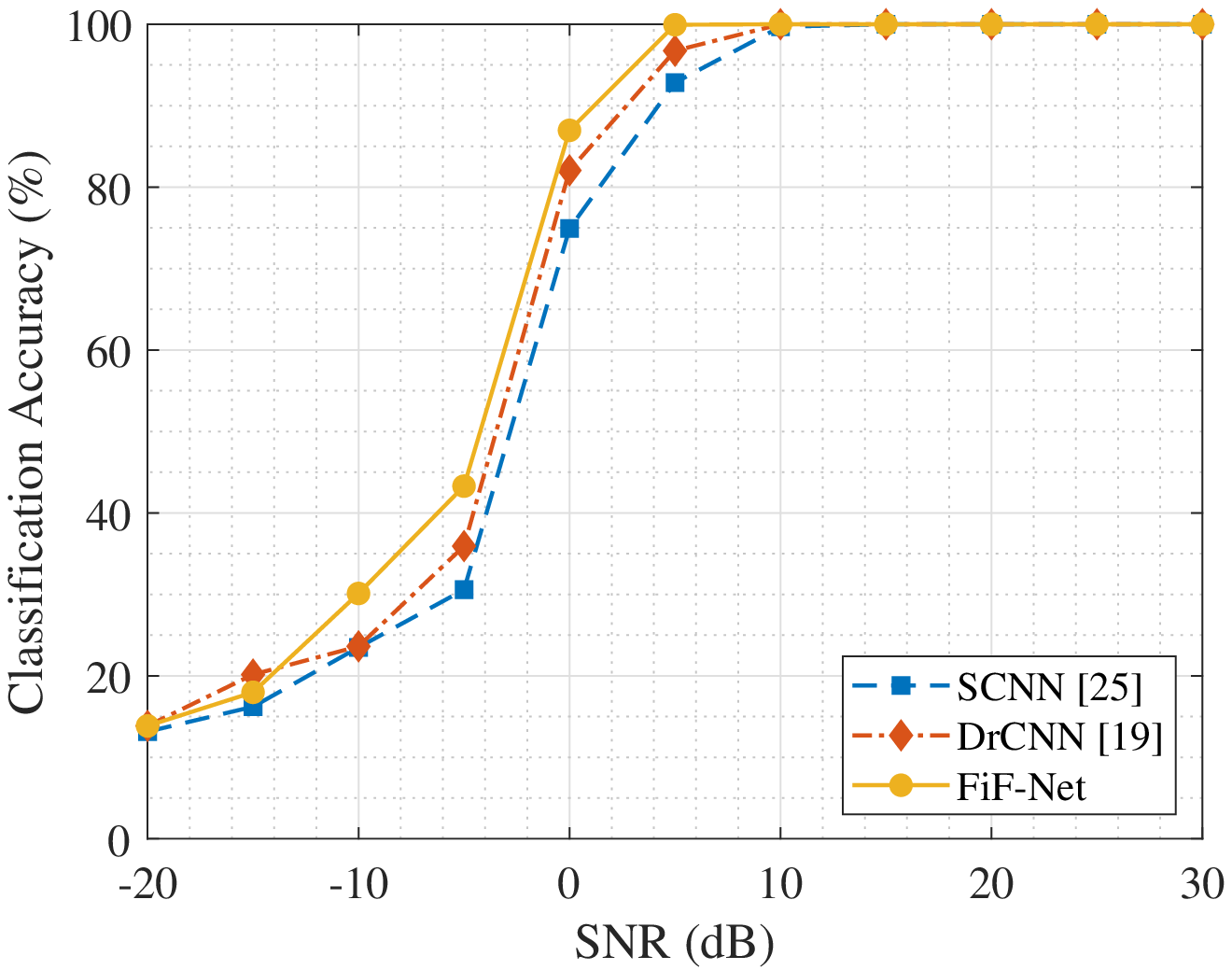}
		\label{fig_comparison}
	}
	\caption{Comparison results: (a) accuracy of FiF-Net with different sizes of constellation image and (b) method comparison in terms of eight-modulation classification accuracy.}
	\label{fig_accuracy_imagetype}
\end{figure}

In the last simulation, we compare FiF-Net with two state-of-the-art CNN models, including DrCNN~\cite{Wang-2019} and SCNN~\cite{Zeng-2019} specified for constellation-based modulation classification.
From the comparison results plotted in Fig.~\ref{fig_comparison}, the proposed CNN outperforms DrCNN and SCNN at the most of SNR levels.
For example, FiF-Net achieves the classification rate of $86.96\%$ at $0$ dB SNR, which is better than DrCNN and SCNN by approximately $4.92\%$ and $12.06\%$, respectively.
Both DrCNN and SCNN have the architecture involving several $3 \times 3$ convolutional layers simply organized in cascade structure, but DrCNN performs more accurately than SCNN by configuring two additional fully connected layers.
However, with the large numbers of hidden nodes specified in two fully connected layers (1024 nodes in the first one and 512 nodes in the second one), DrCNN is heavier than SCNN (just 128 nodes).
Therefore, DrCNN requires more computing resources than SCNN for modulation prediction in the inference process.
By deploying several 1-D asymmetric convolutional layers and one 128-node fully connected layer, FiF-Net can get the good trade-off between accuracy and computational cost.

\section{Conclusion}

In this paper, we have introduced a novel deep convolutional neural network, namely FiF-Net, for constellation-based modulation classification, in which the network architecture is specified by multiple processing blocks to comprehensively learn more intrinsic radio characteristics from constellation diagram.
The deep visual features of constellation points in a diagram are completely learned by deploying grouped and asymmetric convolutional layers concurrently in a flow-in-flow structure.
Based on the performance benchmark with the constellation image dataset of eight digital modulation formats, FiF-Net achieves the classification rate of approximately $87\%$ at $0$ dB SNR under the synthetic channel impairment of multipath Rayleigh fading and AWGN.
Remarkably, with a well design of flow-in-flow structure, FiF-Net outperforms many state-of-the-art deep models for learning modulation patterns from constellation map in terms of accuracy.

%\vspace{12pt}
%\color{red}
%IEEE conference templates contain guidance text for composing and formatting conference papers. Please ensure that all template text is removed from your conference paper prior to submission to the conference. Failure to remove the template text from your paper may result in your paper not being published.

\end{document}